# Molecular Dynamics Simulations Reveal PolyQ-Length-Dependent Conformational Changes in Huntingtin Exon-1: Implications for Environmental Co-Solvent Modulation of Aggregation-Prone States

J. Geddes-Nelson, X. Liu, K.T. Yong*

*School of Biomedical Engineering, Faculty of Engineering, The University of Sydney, Sydney, NSW 2006, Australia*

*Corresponding author: ken.yong@sydney.edu.au

# Abstract

Huntington's disease (HD) is driven by CAG-repeat expansion in the *HTT* gene, encoding polyglutamine (polyQ) tracts that promote huntingtin (HTT) misfolding and aggregation. While experimental studies have characterised polyQ-length-dependent aggregation, the conformational dynamics preceding aggregation remain poorly understood at the atomistic level. Here, we present all-atom molecular dynamics (MD) simulations of HTT exon-1 constructs encompassing the N17 domain, polyQ tracts of three clinically relevant lengths (Q21, wild-type; Q40, adult-onset threshold; Q70, juvenile-onset), and the polyproline (polyP) region. Multi-copy (four-chain) simulations were performed in explicit SPC/E water using the OPLS-AA force field for 100 ns. We quantified radius of gyration (Rg), solvent-accessible surface area (SASA), root-mean-square deviation (RMSD), and intra-protein hydrogen bonds as conformational proxies for aggregation propensity. Expanding polyQ tracts produced progressive increases in Rg (32.1 ± 0.7 Å for Q21 to 40.2 ± 0.5 Å for Q70) and SASA, consistent with adoption of extended, solvent-exposed conformations in pathological variants. Introduction of organic co-solvents (methanol, hexane, trichloroethylene [TCE]) at 0.5–1.0 M concentrations further modulated these conformational landscapes: TCE exposure drove pronounced structural expansion in both Q21 (Rg = 43.9 ± 1.3 Å) and Q40 (Rg = 39.5 ± 0.4 Å) variants, while methanol induced mild compaction in Q21 (Rg = 28.5 ± 0.5 Å). These findings represent, to our knowledge, the first MD investigation of co-solvent effects on HTT exon-1 conformational dynamics. Although single-trajectory limitations preclude definitive mechanistic conclusions, the observed trends suggest that hydrophobic co-solvents may shift the conformational ensemble of HTT exon-1 toward aggregation-prone states, providing a computational framework for investigating gene–environment interactions in HD.

# 1. Introduction

Huntington's disease (HD) is a progressive, autosomal-dominant neurodegenerative disorder caused by an unstable CAG trinucleotide repeat expansion in exon-1 of the huntingtin (*HTT*) gene [1,2]. This expansion encodes an elongated polyglutamine (polyQ) tract within the N-terminal region of the huntingtin protein, and repeat lengths exceeding approximately 36 glutamines are associated with disease pathology, with longer tracts correlating with earlier onset and greater severity [3,4]. The resulting mutant huntingtin (mHTT) protein undergoes misfolding and aggregation, forming intraneuronal inclusion bodies that are a pathological hallmark of HD [5,6].

The N-terminal fragment encoded by HTT exon-1 is both necessary and sufficient for aggregate formation *in vitro* and in cellular models [7,8]. This fragment comprises three functionally distinct domains: the amphipathic N17 region, which mediates membrane association and modulates aggregation kinetics [9,10]; the polyQ tract, whose length determines aggregation propensity; and the C-terminal polyproline (polyP) region, which has been proposed to act as an intrinsic aggregation modulator [11,12]. A downstream HEAT (Huntingtin, Elongation factor 3, protein phosphatase 2A, TOR1) repeat domain further influences the structural context of the exon-1 fragment, though its precise role in aggregation remains debated [13,14].

Molecular dynamics (MD) simulations have emerged as a powerful tool for interrogating the conformational dynamics of polyQ-containing proteins at atomistic resolution. Previous computational studies have revealed that polyQ tracts adopt a heterogeneous ensemble of structures, including collapsed coils, beta-hairpins, and extended conformations, with the population balance shifting toward aggregation-competent states as tract length increases [15–17]. Coarse-grained and all-atom simulations of HTT exon-1 fragments have demonstrated that the N17 domain can adopt alpha-helical conformations that nucleate intermolecular contacts [18,19], while the polyP region may impose steric constraints on polyQ collapse [11]. Recent replica-exchange and multi-microsecond simulations have provided evidence for polyQ-length-dependent changes in interdomain interactions and conformational sampling [20–22].

Despite considerable progress in understanding polyQ-intrinsic determinants of aggregation, substantially less attention has been directed toward environmental modulators of HTT conformational dynamics. Epidemiological evidence suggests that environmental exposures—including organic solvents such as trichloroethylene (TCE)—may influence the onset or progression of neurodegenerative diseases [23,24]. TCE has been linked to increased risk of Parkinson's disease through mechanisms involving mitochondrial complex I inhibition and oxidative stress [25,26]. However, whether organic co-solvents can directly modulate the conformational landscape of aggregation-prone proteins such as mHTT has not been systematically investigated using MD simulations.

In the present study, we employ all-atom MD simulations to characterise the conformational response of HTT exon-1 constructs to (i) polyQ tract length variation across three clinically relevant repeat lengths (Q21, Q40, and Q70), and (ii) exposure to a panel of organic co-solvents (methanol, hexane, acetone, and TCE) representing distinct physicochemical properties. Multi-copy simulations (four protein chains per simulation box) were conducted using the OPLS-AA force field in explicit SPC/E water for 100 ns production runs. Conformational proxies including radius of gyration (Rg), solvent-accessible surface area (SASA), root-mean-square deviation (RMSD), and intra-protein hydrogen-bond counts were computed to assess structural compactness, solvent exposure, and internal stabilisation.

We emphasise that the metrics employed here—Rg, SASA, RMSD, and hydrogen-bond counts—are conformational proxies that report on structural properties associated with, but not equivalent to, aggregation. Direct observation of aggregation events would require substantially longer simulation timescales (multi-microsecond to millisecond) and/or enhanced sampling methods that are beyond the scope of this study [20,27]. Furthermore, all simulations represent single trajectories without independent replicates, which limits statistical inference. Nevertheless, the observed polyQ-length-dependent trends are consistent with experimental observations and prior computational work, and the co-solvent data provide an initial computational framework for exploring gene–environment interactions relevant to HD pathogenesis.

## 2. Methods

### 2.1. Protein Structure Generation

HTT exon-1 constructs were generated using ColabFold v1.5.5 [28], an accelerated implementation of AlphaFold2 [29] utilising MMseqs2-based multiple sequence alignment. Six constructs were modelled: three containing the first HEAT repeat domain (HEAT-containing) and three without (HEAT-less), each at polyQ lengths of Q21 (wild-type), Q40 (adult/late-onset threshold), and Q70 (juvenile/early-onset). Input sequences comprised the N17 amphipathic region, the polyQ tract, the polyproline (polyP) region, and, where applicable, the first HEAT repeat. ColabFold was run with default parameters (Supplementary Table S1) and predicted structures were assessed using pLDDT confidence scores; all constructs showed expected high confidence in structured domains (N17, HEAT repeat) and lower confidence in the intrinsically disordered polyQ and polyP regions.

### 2.2. System Preparation and Simulation Protocol

All MD simulations were performed using GROMACS (versions 2021.4 and 2025.2) [30]. Each simulation system contained four copies of the target protein construct placed in a cubic periodic box with a minimum 1.5 nm buffer from the protein to the box edge. Proteins were parameterised with the OPLS-AA force field [31], and the system was solvated with SPC/E water molecules [32]. System charge was neutralised by adding $Na^+$ and $Cl^-$ counter-ions.

Energy minimisation was performed using the steepest-descent algorithm (5000 steps, convergence threshold of 1000 kJ mol$^{-1}$ nm$^{-1}$ for ion positioning; then 500000 steps with a threshold of 500 kJ mol$^{-1}$ nm$^{-1}$ for the main minimisation). The system was then equilibrated in two stages: (i) NVT equilibration for 1000 ps at the target temperature using the V-rescale thermostat ($\tau_T$ = 0.1 ps) with position restraints on all protein heavy atoms, and (ii) NPT equilibration for 1000 ps using the Berendsen barostat (isotropic coupling, $\tau_P$ = 1.0 ps, compressibility = $4.5 \times 10^{-5}$ bar$^{-1}$) with position restraints maintained.

Production simulations were run for 100 ns with a 2 fs integration timestep. The Parrinello–Rahman barostat [33] replaced the Berendsen barostat for production runs to ensure correct ensemble sampling. Temperature coupling employed the V-rescale thermostat with separate coupling groups for protein and non-protein atoms. Long-range electrostatics were treated with the Particle Mesh Ewald (PME) method [34] (fourth-order interpolation, Fourier spacing of 0.1 nm). Short-range van der Waals and Coulomb interactions used a 1.4 nm cutoff with the Verlet neighbour list scheme. All bond lengths involving hydrogen atoms were constrained using the LINCS algorithm [35] (fourth-order expansion, single iteration). Energy and pressure dispersion corrections were applied. Full simulation parameters are provided in Supplementary Table S2.

## 2.3. Co-Solvent Simulations

To investigate the effect of organic co-solvents on HTT exon-1 conformational dynamics, selected protein constructs (Q21 and Q40, HEAT-less variants) were simulated in the presence of methanol (0.5–1.0 M), hexane (1.0 M), acetone (1.0 M), or trichloroethylene (TCE, 1.0 M). Co-solvent molecules were parameterised using LigParGen [36], which provides OPLS-AA-compatible parameters. Co-solvent molecules were distributed in the simulation box alongside SPC/E water. All other simulation parameters remained identical to the aqueous control simulations described above.

## 2.4. Trajectory Analysis

Trajectory analyses were performed using MDAnalysis [37,38] and GROMACS built-in tools. The following conformational metrics were computed over the full 100 ns trajectory for each simulation:

**Root-mean-square deviation (RMSD):** Backbone RMSD was calculated relative to the energy-minimised starting structure, using least-squares fitting to the initial frame. RMSD reports on global structural drift from the starting conformation.

**Root-mean-square fluctuation (RMSF):** Per-residue RMSF was computed over the production trajectory, averaged across all four protein copies in each simulation. RMSF identifies regions of high conformational flexibility.

**Radius of gyration (Rg):** Rg was calculated as the mass-weighted root-mean-square distance of all atoms from the protein centre of mass. Rg serves as a measure of overall structural compactness; increasing Rg indicates structural expansion.

**Solvent-accessible surface area (SASA):** SASA was computed using a spherical probe of radius 1.4 Å, approximately corresponding to the radius of a water molecule. SASA quantifies the extent of protein surface exposed to solvent, with increases suggesting unfolding or loss of tertiary packing.

**Intra-protein hydrogen bonds:** Hydrogen bonds within each protein chain were enumerated using standard geometric criteria (donor–acceptor distance $\leq 3.0$ Å, donor–H–acceptor angle $\geq 150°$ and distance $\leq 1.2$ Å). Decreasing hydrogen-bond counts may indicate loss of secondary/tertiary structure.

For quantitative comparisons, time-averaged values and standard deviations were computed over the final 20 ns of each trajectory (80–100 ns) to focus on the equilibrated or quasi-equilibrated portion of the simulation. We note that 100 ns may be insufficient for full convergence of all metrics, particularly for the larger Q70 constructs, and this limitation is discussed in Section 5.

## 2.5. Analysis Scope and Limitations

All simulations represent single trajectories (n = 1 per condition) without independent replicates. Consequently, the standard deviations reported reflect temporal fluctuations within a single trajectory, not statistical uncertainty across independent runs. The four protein copies within each simulation box experience the same solvent environment and box-level fluctuations, and therefore do not constitute independent replicates for statistical purposes. This design limits formal statistical comparison between conditions and means that the observed trends should be interpreted as preliminary observations requiring validation through replicate simulations or enhanced sampling approaches.

# 3. Results

## 3.1. PolyQ Length Dependence of Conformational Metrics in Aqueous Controls

To establish baseline conformational behaviour, we first examined the HEAT-less (non-HEAT) variants of HTT exon-1 at the three polyQ lengths in pure aqueous solvent. Figure 1 and Table 1 summarise the key conformational metrics averaged over the final 20 ns (80–100 ns) of each 100 ns trajectory.

A clear polyQ-length-dependent trend was observed for Rg: the wild-type Q21 construct maintained a relatively compact structure (Rg = 32.06 ± 0.69 Å), the Q40 late-onset variant showed a modest increase (Rg = 33.38 ± 0.63 Å), and the Q70 early-onset variant adopted a substantially more expanded conformation (Rg = 40.17 ± 0.50 Å), representing a 25% increase relative to Q21. SASA followed a parallel trend, increasing from 25,705 ± 332 Å$^2$ (Q21) to 29,906 ± 344 Å$^2$ (Q40) to 37,844 ± 369 Å$^2$ (Q70), indicating progressive exposure of protein surface to solvent with increasing polyQ length.

Interestingly, intra-protein hydrogen-bond counts also increased with polyQ length: from 24.0 ± 4.2 (Q21) to 31.2 ± 4.5 (Q40) to 50.3 ± 5.8 (Q70). This likely reflects the greater number of potential hydrogen-bond donors and acceptors in the longer polyQ tracts, combined with the formation of transient intra-chain contacts as the expanded polyQ region samples extended and turn-like conformations. RMSD values were 42.42 ± 0.44 Å (Q21), 33.40 ± 0.67 Å (Q40), and 44.15 ± 1.50 Å (Q70), suggesting substantial structural rearrangement from the AlphaFold2-predicted starting structures in all cases, as expected for intrinsically disordered regions.

These results are consistent with the prevailing model that expanded polyQ tracts adopt increasingly extended, solvent-exposed conformations that may be precursors to aggregation-competent states [15–17]. However, we emphasise that direct aggregation events were not observed on the 100 ns timescale of these simulations.

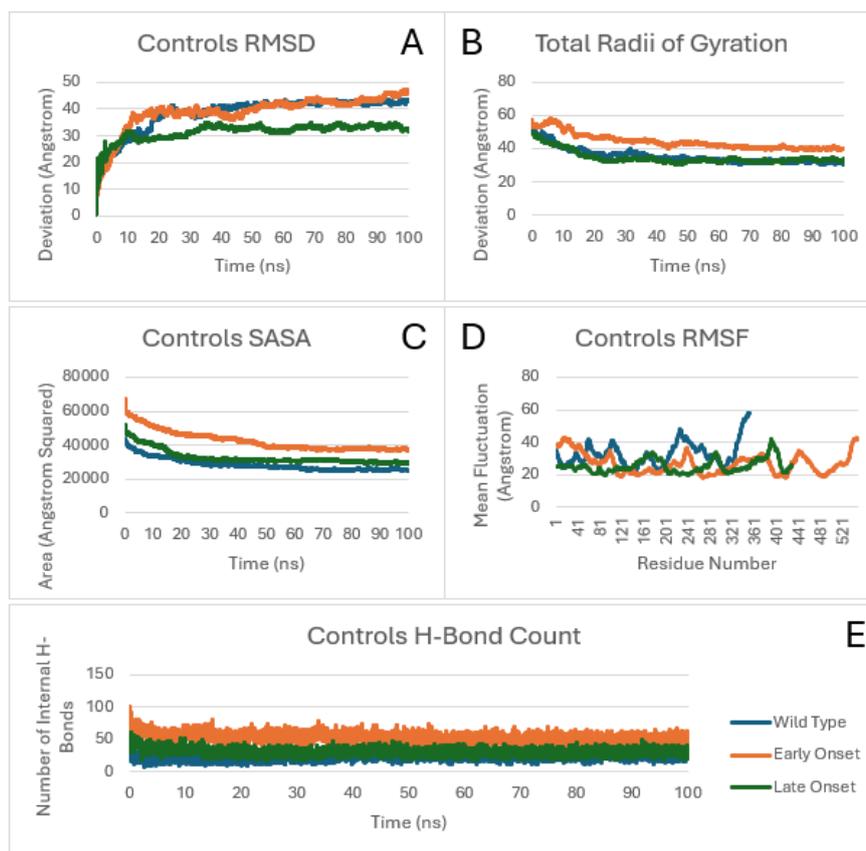

*Figure 1. Analytical summary plots of non-HEAT controls depicting A) Root Mean Square Deviation, B) Total Radii of Gyration, C) Total Solvent Accessible Surface Area, D) Root Mean Square Fluctuation and E) Total Protein H-Bond Count.*

## 3.2. Co-Solvent Effects on Wild-Type (Q21) HTT Exon-1

The wild-type Q21 construct was simulated in the presence of three organic co-solvents to assess whether environmental molecules could perturb the conformational landscape of HTT exon-1. Table 2 and Figure 2 present the results.

Methanol (0.5 M) induced a modest compaction of the Q21 construct, reducing Rg from $32.06 \pm 0.69$ Å (aqueous control) to $28.53 \pm 0.50$ Å, with a concurrent slight decrease in SASA ($25,283 \pm 558$ Å$^2$ vs. $25,705 \pm 332$ Å$^2$). This is consistent with the known kosmotropic-like behaviour of low-concentration methanol, which can stabilise compact protein states [39]. Intra-protein hydrogen bonds decreased to $17.8 \pm 3.4$, potentially reflecting altered solvent–protein hydrogen-bonding competition.

In contrast, hexane (1.0 M) drove structural expansion: Rg increased to 37.10 ± 0.85 Å (+16% relative to aqueous control), and SASA rose to 31,331 ± 578 Å$^2$ (+22%). Hydrogen bonds decreased to 18.7 ± 4.1, suggesting loss of internal stabilisation. As a non-polar, hydrophobic solvent, hexane may preferentially associate with exposed hydrophobic residues, promoting structural unfolding.

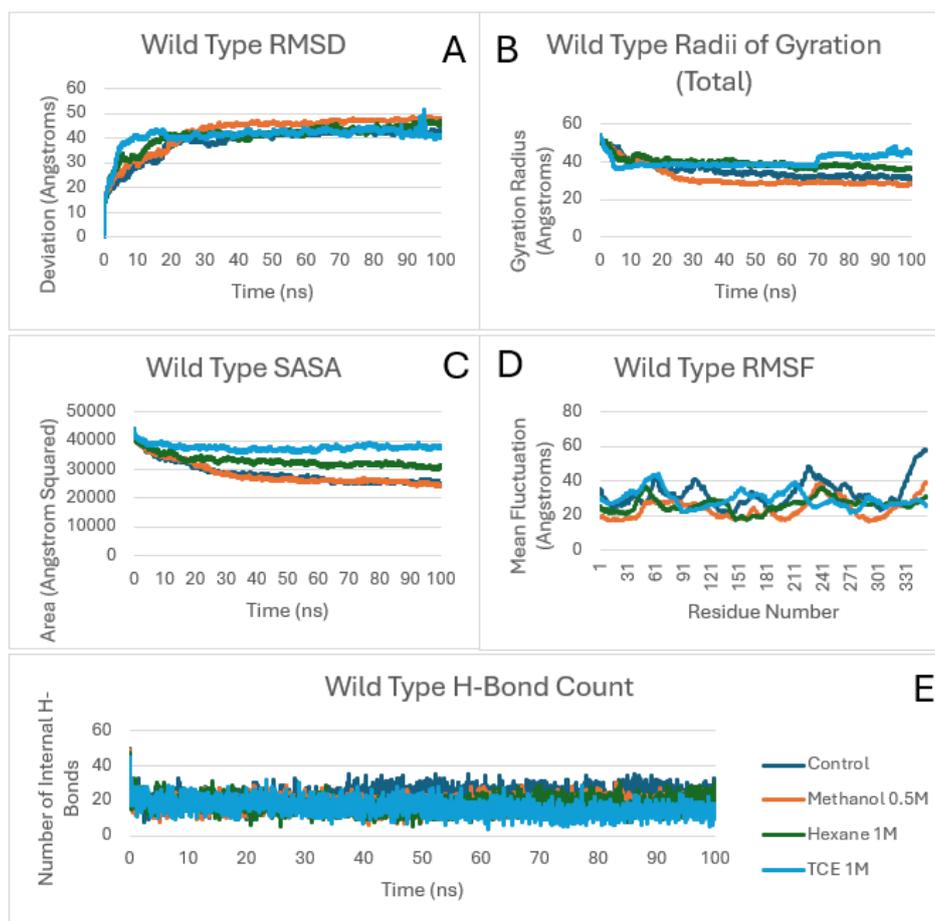

*Figure 2. Analytical summary plots of non-HEAT wt-HTT solvent comparisons depicting A) Root Mean Square Deviation, B) Total Radii of Gyration, C) Total Solvent Accessible Surface Area, D) Root Mean Square Fluctuation and E) Total Protein H-Bond Count.*

The most pronounced effect was observed with TCE (1.0 M), which drove Rg to 43.86 ± 1.26 Å (+37% relative to control), accompanied by a large increase in SASA to 37,729 ± 346 Å$^2$ (+47%) and a dramatic reduction in hydrogen bonds to 13.8 ± 3.4 (–42%). This substantial structural expansion and loss of internal hydrogen bonding in the presence of TCE is notable, as TCE is an environmental toxicant epidemiologically linked to neurodegenerative disease [25,26]. The Q21 + TCE conformational profile (Rg = 43.86 Å) approached that of the pathological Q70 construct in pure water (Rg = 40.17 Å), raising the possibility—albeit speculative at this stage—that certain environmental exposures might shift the conformational ensemble of wild-type HTT toward states resembling those of disease-associated mutants.

## 3.3. Co-Solvent Effects on Late-Onset (Q40) HTT Exon-1

Co-solvent simulations were extended to the Q40 late-onset variant to examine whether polyQ expansion alters susceptibility to co-solvent-induced conformational changes. Table 3 and Figure 3 summarise the results.

Methanol (1.0 M) had minimal effect on Q40 Rg ($32.05 \pm 0.48$ Å vs. $33.38 \pm 0.63$ Å control), with SASA ($29,879 \pm 302$ Å$^2$) and hydrogen bonds ($33.5 \pm 4.9$) remaining comparable to control values. This contrasts with the compaction observed in Q21, suggesting that the expanded polyQ tract may buffer against methanol-induced structural perturbation.

Hexane (1.0 M) produced substantial expansion in Q40, increasing Rg to $40.65 \pm 1.02$ Å (+22% vs. control) and SASA to $37,678 \pm 413$ Å$^2$ (+26%), with hydrogen bonds relatively maintained ($31.0 \pm 4.8$ vs. $31.2 \pm 4.5$ control). This expansion pattern is similar to that observed in Q21 + hexane, suggesting a consistent hydrophobic-solvent-mediated unfolding mechanism.

Acetone (1.0 M) modestly increased Rg in Q40 to $34.01 \pm 0.86$ Å (compared to $33.38 \pm 0.63$ Å control), with a notable increase in hydrogen bonds to $47.6 \pm 6.2$, potentially indicating formation of new intra-chain contacts facilitated by acetone's dual hydrophilic/hydrophobic character.

TCE (1.0 M) increased Q40 Rg to $39.50 \pm 0.35$ Å (+18%), with the most dramatic change observed in SASA, which rose to $43,023 \pm 481$ Å$^2$ (+44% relative to control). Hydrogen bonds increased to $42.0 \pm 5.6$, potentially reflecting TCE-induced reorganisation that exposes the polyQ tract while promoting new backbone contacts. This complex response pattern differs from the simpler expansion-plus-hydrogen-bond-loss observed in Q21 + TCE, suggesting that the longer polyQ tract modulates the protein's response to hydrophobic co-solvents.

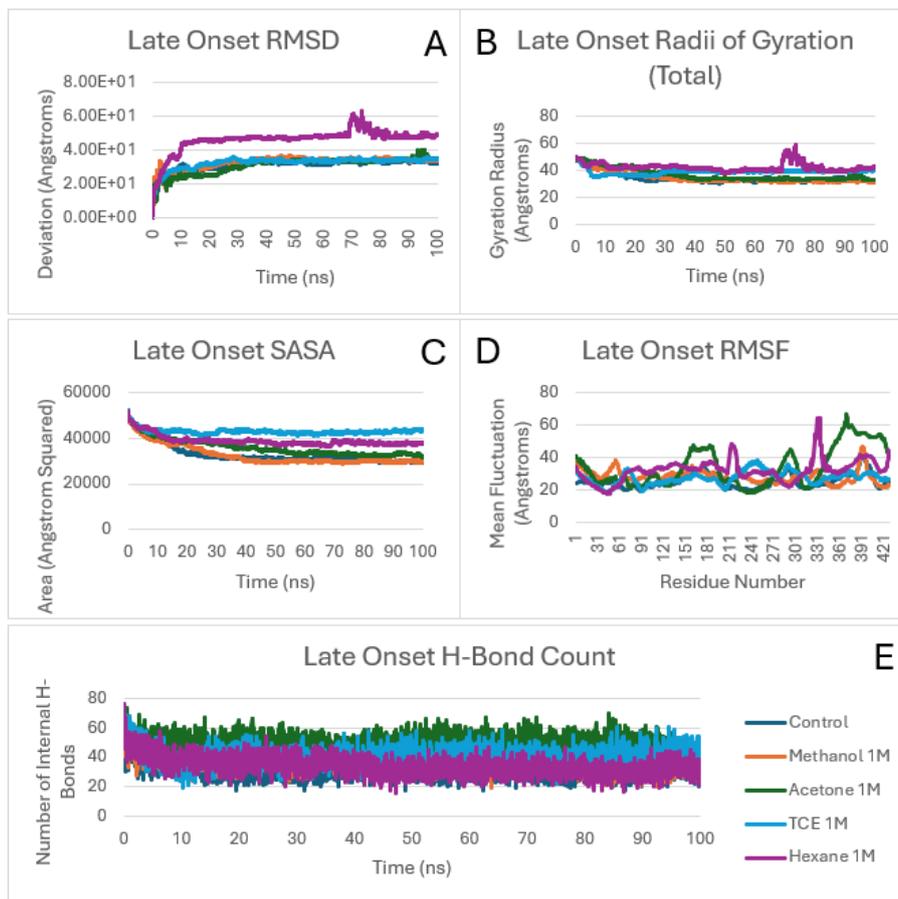

*Figure 3. Analytical summary plots of non-HEAT lo-HTT solvent comparisons depicting A) Root Mean Square Deviation, B) Total Radii of Gyration, C) Total Solvent Accessible Surface Area, D) Root Mean Square Fluctuation and E) Total Protein H-Bond Count.*

## 3.4. Summary of Conformational Trends

Across all conditions examined, two principal trends emerged: (i) increasing polyQ length drives progressive structural expansion, as measured by Rg and SASA, under aqueous conditions; and (ii) hydrophobic co-solvents, particularly TCE and hexane, amplify this expansion, with the magnitude of the effect depending on both the co-solvent identity and the polyQ length. Methanol exhibited a qualitatively different behaviour, inducing compaction in Q21 but having minimal effect on Q40. These observations are summarised schematically in Table 4. We note that all trends are based on single-trajectory simulations and require validation through replicate studies.

**Table 1. Conformational metrics for aqueous control simulations (HEAT-less variants), averaged over the final 20 ns.**

| Construct | PolyQ Length | RMSD (Å) | Rg (Å) | SASA (Å²) | H-Bonds |
|-----------|--------------|----------|--------|-----------|---------|
| Wild-type | Q21 | 42.42 ± 0.44 | 32.06 ± 0.69 | 25,705 ± 332 | 24.0 ± 4.2 |
| Late-onset | Q40 | 33.40 ± 0.67 | 33.38 ± 0.63 | 29,906 ± 344 | 31.2 ± 4.5 |
| Early-onset | Q70 | 44.15 ± 1.50 | 40.17 ± 0.50 | 37,844 ± 369 | 50.3 ± 5.8 |

**Table 2. Co-solvent effects on wild-type (Q21) HTT exon-1 conformational metrics (HEAT-less variants).**

| Condition | RMSD (Å) | Rg (Å) | ΔRg (%) | SASA (Å²) | H-Bonds |
|-----------|----------|--------|---------|-----------|---------|
| Aqueous control | 42.42 ± 0.44 | 32.06 ± 0.69 | — | 25,705 ± 332 | 24.0 ± 4.2 |
| + Methanol (0.5 M) | 47.61 ± 0.43 | 28.53 ± 0.50 | −11% | 25,283 ± 558 | 17.8 ± 3.4 |
| + Hexane (1.0 M) | 45.03 ± 0.95 | 37.10 ± 0.85 | +16% | 31,331 ± 578 | 18.7 ± 4.1 |
| + TCE (1.0 M) | 42.24 ± 1.43 | 43.86 ± 1.26 | +37% | 37,729 ± 346 | 13.8 ± 3.4 |

**Table 3. Co-solvent effects on late-onset (Q40) HTT exon-1 conformational metrics (HEAT-less variants).**

| Condition | RMSD (Å) | Rg (Å) | ΔRg (%) | SASA (Å²) | H-Bonds |
|-----------|----------|--------|---------|-----------|---------|
| Aqueous control | 33.40 ± 0.67 | 33.38 ± 0.63 | — | 29,906 ± 344 | 31.2 ± 4.5 |
| + Methanol (1.0 M) | 34.60 ± 0.70 | 32.05 ± 0.48 | −4% | 29,879 ± 302 | 33.5 ± 4.9 |
| + Hexane (1.0 M) | 48.42 ± 0.87 | 40.65 ± 1.02 | +22% | 37,678 ± 413 | 31.0 ± 4.8 |
| + Acetone (1.0 M) | 34.76 ± 1.55 | 34.01 ± 0.86 | +2% | 32,756 ± 451 | 47.6 ± 6.2 |
| + TCE (1.0 M) | 34.51 ± 0.51 | 39.50 ± 0.35 | +18% | 43,023 ± 481 | 42.0 ± 5.6 |

# 4. Discussion

## 4.1. PolyQ-Length-Dependent Conformational Expansion

The progressive increase in Rg and SASA with polyQ length observed in our aqueous control simulations is broadly consistent with the biophysical understanding of polyQ expansion in HD. Experimental studies using techniques such as fluorescence correlation spectroscopy [4], small-angle X-ray scattering (SAXS) [40], and atomic force microscopy (AFM) [6] have demonstrated that longer polyQ tracts adopt more extended conformations and exhibit greater aggregation propensity. Our simulations recapitulate this length-dependent expansion, with Q70 showing a 25% increase in Rg relative to Q21.

The observed trend is also consistent with prior MD studies. Jakubek et al. [15] reported that polyQ chains of increasing length show progressively greater structural heterogeneity and propensity for extended states. Kang et al. [22] used temperature replica-exchange MD (T-REMD) to demonstrate that expanded polyQ tracts sample a broader conformational landscape with increased populations of beta-sheet-like structures. Mohanty et al. [20,21] performed multi-microsecond aggregate simulations of HTT exon-1 and observed polyQ-length-dependent changes in interdomain interactions and compactness. While our 100 ns simulations are considerably shorter than these benchmark studies, the qualitative agreement in polyQ-length-dependent expansion trends provides confidence that our simulations capture fundamental aspects of the conformational response.

We note that the absolute RMSD values (33–44 Å) are large relative to typical globular protein simulations, reflecting the intrinsically disordered nature of the HTT exon-1 polyQ and polyP regions. Large RMSD values are expected when the starting structure is an AlphaFold2 prediction of an intrinsically disordered protein, as the simulation rapidly departs from the static predicted conformation to explore the accessible conformational ensemble. Accordingly, RMSD should not be interpreted as a measure of instability but rather as confirmation that the simulated protein samples a broad conformational space, consistent with its disordered character.

## 4.2. Co-Solvent Modulation of Conformational Landscapes

The co-solvent results represent, to our knowledge, the first systematic MD investigation of organic co-solvent effects on HTT exon-1 conformational dynamics. The differential responses observed across co-solvents and polyQ lengths suggest that protein–co-solvent interactions are governed by a complex interplay of hydrophobic effects, hydrogen-bonding competition, and preferential solvation.

Methanol exhibited a mild compaction effect on Q21 (Rg decreased by 11%) but had negligible impact on Q40. This is consistent with the known behaviour of short-chain alcohols at low concentrations, where they can act as kosmotropes that preferentially hydrate the protein surface and stabilise compact states [39]. The attenuated response in Q40 may reflect the greater conformational entropy of the expanded polyQ tract, which resists compaction.

Hexane consistently promoted structural expansion in both Q21 (+16% Rg) and Q40 (+22% Rg), accompanied by SASA increases. As a purely hydrophobic solvent, hexane likely disrupts the hydrophobic core interactions that maintain compact conformations, while being unable to compete for hydrogen bonds with the protein backbone. The decrease in intra-protein hydrogen bonds observed in Q21 + hexane (18.7 vs. 24.0 in control) supports a mechanism involving disruption of internal packing.

TCE produced the most dramatic effects across both polyQ lengths. In Q21, TCE drove a 37% increase in Rg, a 47% increase in SASA, and a 42% decrease in hydrogen bonds, suggesting a profound loss of structural compactness and internal stabilisation. In Q40, TCE increased Rg by 18% and SASA by 44%, though hydrogen bonds increased rather than decreased. This divergent hydrogen-bond response is intriguing and may reflect polyQ-length-dependent differences in the mode of TCE interaction: in Q21, TCE may primarily disrupt existing contacts, while in Q40, the reorganisation of the larger polyQ tract may create new intra-chain hydrogen-bonding opportunities.

The observation that TCE exposure shifts the conformational profile of wild-type Q21 toward metrics resembling the pathological Q70 construct is provocative, though we emphasise that this comparison is based on a single conformational metric ($R_g$) from single-trajectory simulations. The epidemiological association between TCE exposure and neurodegenerative disease risk [25,26] provides environmental relevance to this observation, but establishing a causal molecular mechanism would require substantially more extensive computational and experimental investigation.

### 4.3. Mechanistic Considerations

Based on the observed trends, we propose a tentative classification of co-solvent effects on HTT exon-1 conformational dynamics:

**Polar protic solvents (methanol):** May stabilise compact states at low concentrations through preferential hydration, but with diminished effect on expanded polyQ tracts.

**Non-polar solvents (hexane):** Promote structural expansion through disruption of hydrophobic packing, with consistent effects across polyQ lengths.

**Chlorinated solvents (TCE):** Drive the most pronounced expansion and conformational reorganisation, potentially through a combination of hydrophobic disruption and specific interactions with the polyQ backbone. The magnitude of TCE effects may reflect its intermediate polarity and ability to penetrate both hydrophobic and polar regions of the protein.

**Polar aprotic solvents (acetone):** Show modest effects on global metrics but may promote intra-chain hydrogen-bond formation, suggesting a distinct mode of interaction.

We emphasise that these classifications are tentative and based on single-trajectory simulations without replicates. The proposed mechanisms should be tested through (i) replicate simulations with independent initial conditions, (ii) analyses of preferential solvation coefficients, (iii) per-residue decomposition of co-solvent contacts, and (iv) free-energy calculations.

## 4.4. Comparison with Literature

Our aqueous control results are qualitatively consistent with the broader HTT MD simulation literature. Moldovean and Chiş [16] reviewed computational studies of HTT aggregation and highlighted the importance of polyQ-length-dependent conformational sampling. Kang et al. [22] reported that longer polyQ tracts show increased beta-sheet propensity in T-REMD simulations, a structural feature associated with aggregation nucleation. While our standard MD simulations at 100 ns timescale do not capture beta-sheet formation dynamics directly, the observed expansion and increased solvent exposure are consistent with a population shift toward less compact, more aggregation-prone states.

The co-solvent simulations extend the existing literature into new territory. While cosolvent MD is widely used for protein engineering and binding-site mapping [39], its application to neurodegenerative disease-relevant proteins is novel. The closest comparable work involves studies of osmolyte effects on polyQ peptides [17], which showed that protective osmolytes like TMAO can stabilise compact polyQ conformations. Our observation that hydrophobic co-solvents destabilise compact states is mechanistically complementary, suggesting that the conformational equilibrium of polyQ proteins is sensitive to the chemical nature of the surrounding solvent environment.

## 5. Limitations

Several important limitations of this study must be acknowledged.

**Single-trajectory design:** All simulations represent single trajectories (n = 1 per condition). The reported standard deviations reflect temporal fluctuations within individual trajectories, not statistical uncertainty across independent replicate simulations. This fundamentally limits formal statistical comparison between conditions and means that the observed trends should be considered preliminary observations rather than statistically validated conclusions. Future work should include a minimum of three to five independent replicate simulations per condition with different initial velocity assignments.

**Limited simulation timescale:** Production runs of 100 ns are short relative to the timescales required for polyQ aggregation (microseconds to milliseconds) and even for full conformational sampling of intrinsically disordered proteins. The metrics reported here capture early-stage conformational dynamics and quasi-equilibrium properties within accessible simulation windows, but may not reflect the true thermodynamic equilibrium ensemble. Convergence of Rg and SASA was assessed visually from time-series plots (Supplementary Figures), and while most metrics appeared to plateau within the analysis window, formal convergence tests (e.g., block averaging or autocorrelation analysis) were not performed.

**Four-copy system design:** The use of four protein copies per simulation box was intended to sample multiple conformational trajectories simultaneously. However, these copies share the same solvent environment and periodic boundary conditions and therefore are not independent. Any inter-copy interactions (e.g., transient contacts) would conflate single-chain conformational dynamics with inter-molecular association, though no persistent inter-copy contacts were observed during visual inspection of trajectories.

**Force field limitations:** The OPLS-AA force field, while well-established, has known limitations for intrinsically disordered proteins and may overestimate secondary structure propensity or underestimate the sampling of extended conformations [31]. More recently developed force fields (e.g., CHARMM36m, a99SB-disp) with improved IDP parameterisation may yield quantitatively different results. Force field choice is a significant source of systematic uncertainty in polyQ simulations [16].

**AlphaFold2-predicted starting structures:** Starting structures were generated using ColabFold/AlphaFold2, which predicts single static structures with limited accuracy for intrinsically disordered regions. While the simulations rapidly departed from these starting structures (as evidenced by large RMSD values), the choice of initial conformation may influence the accessible conformational landscape within a 100 ns trajectory due to kinetic trapping.

**Co-solvent parameterisation:** Organic co-solvents were parameterised using LigParGen with OPLS-AA parameters. While this ensures compatibility with the protein force field, the accuracy of these parameters for liquid mixture simulations at the concentrations employed has not been independently validated. The concentrations used (0.5–1.0 M) are higher than typical environmental exposures, and the results should be interpreted in terms of conformational sensitivity rather than physiological relevance.

**Analysis scope:** The conformational metrics employed (Rg, SASA, RMSD, hydrogen bonds) are established proxies for protein structural properties but do not directly report on aggregation. No inter-chain contact analysis, secondary structure assignment, or free-energy calculations were performed in this study. Direct assessment of aggregation propensity would require analysis of intermolecular contacts, nucleation events, and substantially longer simulation timescales.

**GROMACS version consistency:** Simulations were performed using both GROMACS 2021.4 and 2025.2. While the core algorithms are consistent between versions, minor numerical differences in implementation could contribute to small variations between simulation sets.

**Incomplete early-onset co-solvent data:** Co-solvent simulations for the Q70 (early-onset) variant were not completed in this study, limiting the assessment of co-solvent effects across the full range of clinically relevant polyQ lengths.

## 6. Future Directions

The preliminary observations reported here suggest several productive avenues for future investigation. First, replicate simulations ($n \geq 3$–5 per condition) with independently generated initial velocities are essential to establish statistical confidence in the observed trends and enable formal hypothesis testing. Second, enhanced sampling methods such as replica-exchange MD (REMD), metadynamics, or adaptive sampling algorithms would enable more thorough exploration of the conformational free-energy landscape on computationally accessible timescales. Third, extension of co-solvent simulations to the Q70 early-onset variant would complete the polyQ-length series and test whether the dramatic TCE-induced expansion observed in Q21 is amplified or attenuated by further polyQ expansion.

Additional analyses derivable from the existing trajectories—including per-residue RMSF decomposition, secondary structure assignment (DSSP), principal component analysis (PCA), inter-chain contact frequency maps, and preferential solvation coefficient calculations—would provide deeper mechanistic insight and are planned for subsequent publications. Free-energy perturbation or thermodynamic integration calculations could quantify the energetic basis of co-solvent effects on polyQ conformational equilibria.

Finally, experimental validation of the computational predictions through biophysical techniques such as circular dichroism (CD), dynamic light scattering (DLS), or single-molecule FRET would strengthen the translational relevance of this work and establish whether the in silico observations reflect experimentally measurable phenomena.

# 7. Conclusions

We have presented all-atom molecular dynamics simulations of huntingtin exon-1 constructs spanning three clinically relevant polyglutamine lengths (Q21, Q40, Q70) in aqueous solution and in the presence of organic co-solvents. The simulations reveal a clear polyQ-length-dependent trend toward structural expansion and increased solvent exposure, consistent with the biophysical basis of polyQ-driven aggregation in Huntington's disease. Introduction of organic co-solvents, particularly trichloroethylene (TCE) and hexane, amplified this structural expansion, while methanol induced compaction in the wild-type Q21 variant.

These findings provide an initial computational framework for investigating how environmental co-solvents may modulate the conformational dynamics of aggregation-prone proteins. The observation that TCE shifts wild-type HTT exon-1 conformational metrics toward values resembling those of pathological polyQ-expanded variants, while speculative and requiring validation, raises interesting questions about potential gene–environment interactions in HD pathogenesis.

We emphasise that these results are derived from single-trajectory simulations of limited timescale and should be interpreted as hypothesis-generating observations rather than definitive mechanistic conclusions. The conformational metrics employed are proxies for aggregation propensity, not direct measures of aggregation itself. Nonetheless, the consistent trends across polyQ lengths and co-solvent conditions provide a foundation for more rigorous computational and experimental investigation of environmental modulators of HTT misfolding.

# Figure Legends

**Figure 1.** Conformational metrics for aqueous control simulations of HEAT-less HTT exon-1 variants. Time-series plots (0–100 ns) for backbone RMSD (Å), RMSF (per-residue, averaged across four chains), radius of gyration (Rg, Å), SASA ($Å^2$), and intra-protein hydrogen-bond counts for Q21 (wild-type), Q40 (late-onset), and Q70 (early-onset) constructs.

**Figure 2.** Co-solvent effects on wild-type (Q21) HTT exon-1 conformational dynamics. Time-series plots as in Figure 1 for Q21 in the presence of methanol (0.5 M), hexane (1.0 M), and TCE (1.0 M).

**Figure 3.** Co-solvent effects on late-onset (Q40) HTT exon-1 conformational dynamics. Time-series plots as in Figure 1 for Q40 in the presence of methanol (1.0 M), hexane (1.0 M), acetone (1.0 M), and TCE (1.0 M).

**Supplementary Figure S1–S6.** ColabFold-predicted structures and pLDDT confidence scores for all six HTT exon-1 constructs (HEAT-containing and HEAT-less variants at Q21, Q40, Q70).